\documentclass[10pt]{iopart}

\usepackage{iopams}  
\usepackage{graphicx}
\usepackage[usenames,dvipsnames]{color}
\usepackage{subfig}
\usepackage{color}
\usepackage{float}
\usepackage{subfloat}
\usepackage{caption}
\usepackage{dcolumn}
\usepackage{bm}
\usepackage{multirow}
\usepackage{url}
\usepackage{setspace}
\usepackage{dsfont}

\newcommand{\cmmnt}[1]{\ignorespaces}

\begin{document}

\title[]{Heuristic data analysis for photon detection in single shot RIXS at a Free Electron Laser}

\author{Simone Laterza$^1$, Antonio Caretta$^2$, Barbara Casarin$^1$, Roberta Ciprian$^2$, Martina Dell'Angela$^3$, Wilfried Wurth$^4$, Fulvio Parmigiani$^1$ and Marco Malvestuto$^{1,2}$}

\address{$^1$ Department of Physics, University of Trieste, Via A. Valerio 2, 34127 Trieste, Italy}
\address{$^2$ Elettra-Sincrotrone Trieste S.C.p.A. Strada Statale 14 - km 163.5 in AREA Science Park 34149 Basovizza, Trieste, Italy}
\address{$^3$ CNR-IOM, Strada Statale 14 - km 163.5, 34149 Trieste, Italy}
\address{$^4$ Physics Department and Center for Free-Electron Laser Science, Hamburg University, 22607 Hamburg, Germany}
\ead{marco.malvestuto@elettra.eu}

\date{\today}
\begin{abstract}
A heuristic method based on an image recognition approach has been developed for fast and efficient processing of single shot FEL RIXS spectra. 
\end{abstract}
\maketitle
\ioptwocol

\section{Introduction}

Resonant inelastic X-ray scattering (RIXS) is an established powerful experimental technique for studying the low-energy lattice, charge, orbital and spin excitations in strongly correlated electron and magnetic systems, by measuring the energy momentum and polarization dependence of scattered photons. \cite{Ament:2011jy,Chiuzbaian:2005if,Chiuzbaian:2008jb,Ghiringhelli:2005kp,Wray:2015htb,Schlappa:2012hj}
The possibility of extending this spectroscopy to the time domain with pulsed coherent extreme ultra-violet (EUV) and X-ray Free electron lasers (FELs) promises to disclose an in depth knowledge about the out-of-equilibrium dynamics of the excited states in condensed matter. \cite{DellAngela:2013ve,DellAngela:2016ua,Beye:2010ii,Beye:2013hh}
 
In this respect, RIXS can be employed to map out the temporal evolution of low-energy excitations in the sub-picosecond timescale.

Conventional RIXS experiments carried out at synchrotrons are based on an integration method \cite{Kummer:2017bw}, where the final RIXS spectrum is obtained by collecting a reasonable number of scattered photons from a continuous x-ray source until the desired signal to noise ratio (S/N) is obtained.

On the contrary, by exploiting the pulsed structure of the FEL source, single shot FEL RIXS spectra can be measured, which typically consist of 2D images made of a limited number of collected photons owing to the very small yield of the fluorescence process (~10$^{-2}$\% at the M-edges and ~1\% at L edges of typical transition metals)\cite{Krause:1979fx,Physics:ku,clark1963encyclopedia}.

While a large collection of consecutive FEL single shot RIXS images is necessarily needed to get an integrated RIXS spectrum with an acceptable S/N ratio \cite{DellAngela:2016ua}, on the other hand the analysis of distinct single shot RIXS spectra made of a small number of scattered photons can disclose the opportunity to investigate non linear and coherence effects of the photon-matter interaction process, a possibility that is precluded in continuous synchrotron sources.

A clear drawback of this multi-hit FEL acquisition scheme is the large number of single shot data generated, where few photons per shot are collected together with unnecessary data e.g. background noises, spurious events, cosmic ray events etc.
This require an efficient, reliable and fast extraction method capable of retrieving the key parameters of the collected photons and discarding unnecessary data.

In this paper we report on a heuristic method that has been adopted to retrieve the key parameters of the elastic and inelastic scattered photons collected at each FEL pulse from the probed sample by the RIXS spectrometer.
This method is based on an image recognition approach. 
In brief, the scattered photons that are collected by the spectrometer are imaged by a single photon counting multi-hit solid state detector on a FEL shot-to-shot basis. 
Each raw data image is then processed for extracting the number of collected photons per shot, together with their position on the detector and their energy loss with respect to the incoming photon energy. 
These information are finally used to reconstruct a traditional RIXS spectrum.
  
\begin{figure}[ht!]
\captionsetup{justification=centerlast}
\centering
\includegraphics[width=1\columnwidth]{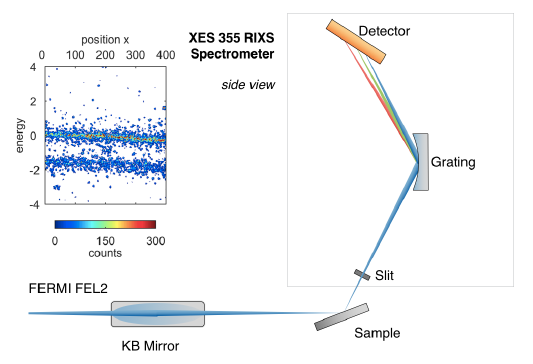}
\caption{\label{exp_setup}\small{Optical layout of the experimental setup. The FEL beam is focused on the sample by a Kirkpatrick–Baez (KB) refocusing mirror system. The photon spectrometer collects the scattered light through a pair of collimating slits and disperses the photons over the solid state detector. A typical unprocessed Cu M$_3$ RIXS 2D image taken from a single crystal CuGeO$_3$ and integrated over 50000 FEL shots ($\sim$1000 seconds at 50 Hz FEL repetition rate) is also displayed. Elastic and inelastic photons distribute along iso-energetic lines.}}

\end{figure}
\section{Experimental setup and data acquisition}

A setup for FEL RIXS experiments is available at the FERMI FEL (ref http), which is capable of operating at a repetition rate of 50 Hz and in the EUV (40-300 eV) energy range. Typical photon fluences at the sample are within the 0.01-10 microJoule/cm$^2$ range.
The X-ray optical layout of the experimental setup is sketched in figure \ref{exp_setup} and it has been already reported in a previous publication \cite{DellAngela:2016ua}.

The scattering plane was the vertical plane of polarization of the incoming photons with the spectrometer at 90$^{\circ}$ to the beam, while the pulses impinged at an angle of 75$^{\circ}$ to the surface.

The FEL photons scattered from the sample were collected by a variable line spacing (VLS) grating Scienta XES355 spectrometer \cite{Nordgren:2000jja}, where the photons are collimated by a set of entrance slits on a spherical grating working in energy dispersive mode. As a consequence, the energy of the photons directly corresponds to the vertical positions on the image.
The energy resolved photons are detected by a 2D position sensitive single photon counting, multi-hit-capable detector. This detector is made of a micro-channel plate (MCP) coupled to a phosphorus screen which is finally imaged by a CMOS camera.
A detected photon triggers an avalanche electric current in the MCP, which in turn generates a luminous spark on the phosphorous screen. 
This detection scheme exhibits a quantum efficiency (defined as the ratio between the number of revealed and incoming photons) of about 20\%. 

Single shot RIXS 2D (400x1100 pixels) images are recorded by a Basler\cite{BaslerAce:wd} camera conveniently synchronized to the timing and data storage systems of FERMI and are encoded as a matrices of binary values.

In this work, reference copper M$_3$-edge ($h\nu=$ 62 eV) RIXS data from a single crystal of CuGeO$_3$ are presented. 
The spectra have been acquired at 50 Hz FEL repetition rate. 
The photon fluence was limited to 11 $\mu$J/cm$^2$ in order to limit sample damaging. A low angle incidence geometry ($\sim$75$^{\circ}$) with respect to the surface normal was adopted.

Data analysis has been made on h5 files, each of them containing 500 images with 400 x 1100 pixel size, totalizing roughly 50000 images per run (1000 seconds acquisition time). A typical unprocessed RIXS 2D image integrated over a run is displayed in Fig. \ref{exp_setup}, where the elastic photons accumulate around the zero energy loss. The image on the detector consists of isoenergetic lines along one direction and the energy dispersion direction perpendicular to these lines. A total amount of around 340000 single shots were recorded during this experiment. The average frequency of recorded events was about 0.06 per image, but clusters did also appear. The average time to process an image was  turned out to be roughly 0.009s.

In order to gain insight into the performance of the data analysis, a simulation of test images was performed. The performance is affected both on the frequency of events on each image and on its size. Except a smaller time on eventless images, average time has a linear dependency on the frequency, whereas its linear dependency on the size of the image is due to the indicization of the elements of images upon which data analysis evaluates whether specified conditions  are met. In order to take into account both data analysis and data loading, the test was performed on a set of 10 specimen files each of them containing 500 computer simulated images of noise and events. The files have been processed 10 times in order to take the average on the runs.
\section{Data processing}

The data processing of single shot FEL-RIXS images consists of a heuristic photon discovery and extraction method \cite{Chan:2012ek}, which aims at extracting from each single shot image the number of photons per image and their absolute pixel positions. 
This single photon counting approach is appropriate because of the very low impinging rate on the sensor. This is a common situation when the RIXS acquisition rate follows the pulse structure of the FEL source.
Since photons are detected as luminous clouds (herein named "blobs"), their energy-pixel position is defined by the center of mass of the cloud. 

\begin{figure}[!h]
\captionsetup{justification=centerlast}
\centering
\includegraphics[width=0.8\columnwidth]{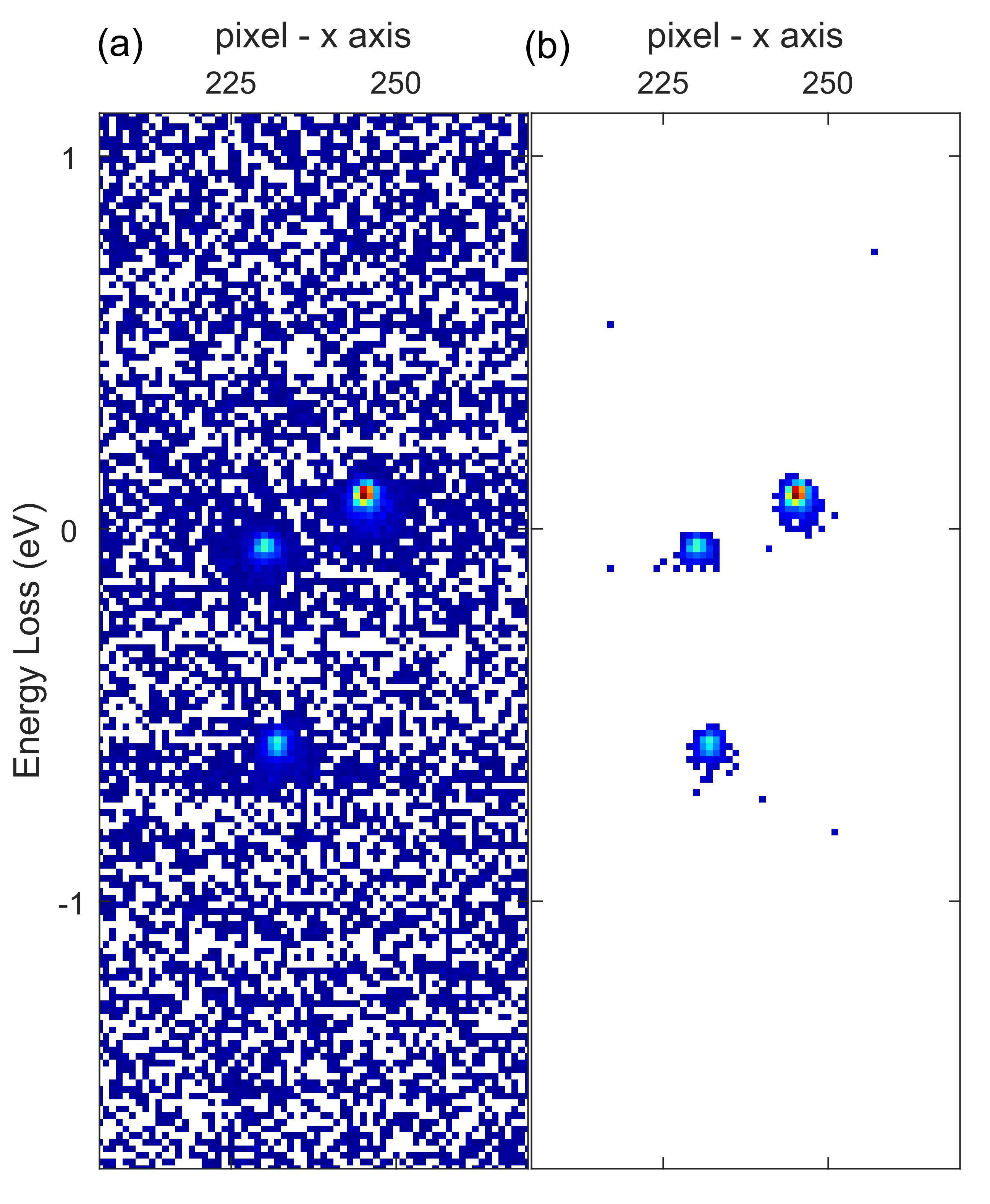}
\begin{singlespace}
\caption{\label{fig:first_image.png} (a): a close-up region of a single shot raw RIXS 2D image. The photon energy loss is resolved along the y-axis by the photon spectrometer. (b) A de-noised image made by subtracting a noise threshold level to the initial raw image. Noise reduction allows also for an initial segmentation of the digital image.}
\end{singlespace}
\end{figure}

A representative single-shot FEL RIXS raw image is shown in Fig. \ref{fig:first_image.png}a, where typical photon footprints are clearly visible. Photon traces show up as round-shaped spots consisting of an assembly of pixels (macropixels) of the single shot RIXS image.
The pixels in the y-axis are directly converted in energy loss values in eV. It can be noted that the photon luminous clouds are overlaid to the statistical and electronic noise.  

The blob discovery and extraction method, which will be discussed in the following paragraphs, is based on the following sequential actions: image noise reduction and segmentation; macropixel labelling and deblurring.

The method described in this paper has been implemeted in a Python \cite{python:ud} code suite, which is available as Github package\cite{OpenSour}.

\begin{figure}[!h]
\captionsetup{justification=centerlast}
\begin{center}

\includegraphics[width=1\columnwidth]{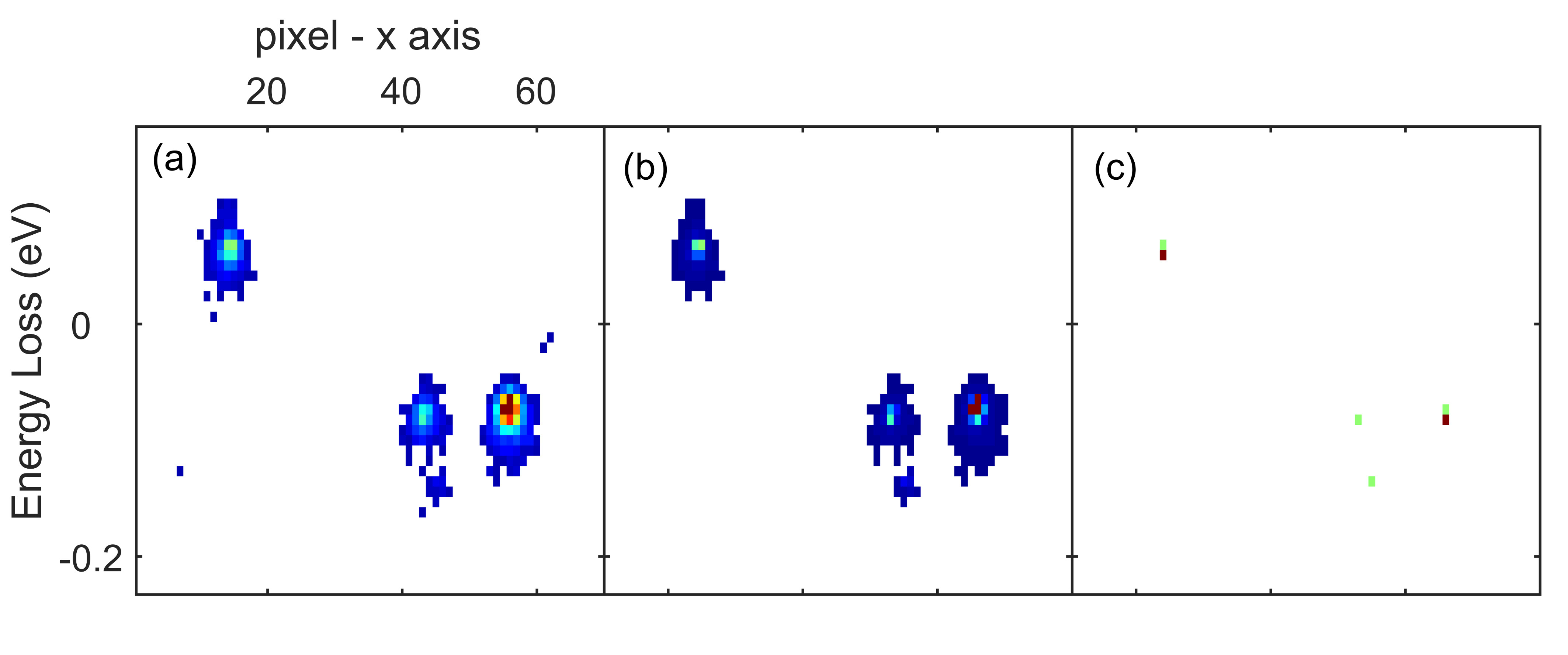}
\begin{singlespace}
\caption[Diagram.]{\label{fig:seventh_image.png}\small{(a) distorted photon blobs. (b) deblurred photon traces. (c) comparison between the centre of masses of distorted and deblurred photon traces.}}
\end{singlespace}

\end{center}

\end{figure}
\begin{figure*}[!htb]
\captionsetup{justification=centerlast}
\begin{center}

\includegraphics[width=1.8\columnwidth]{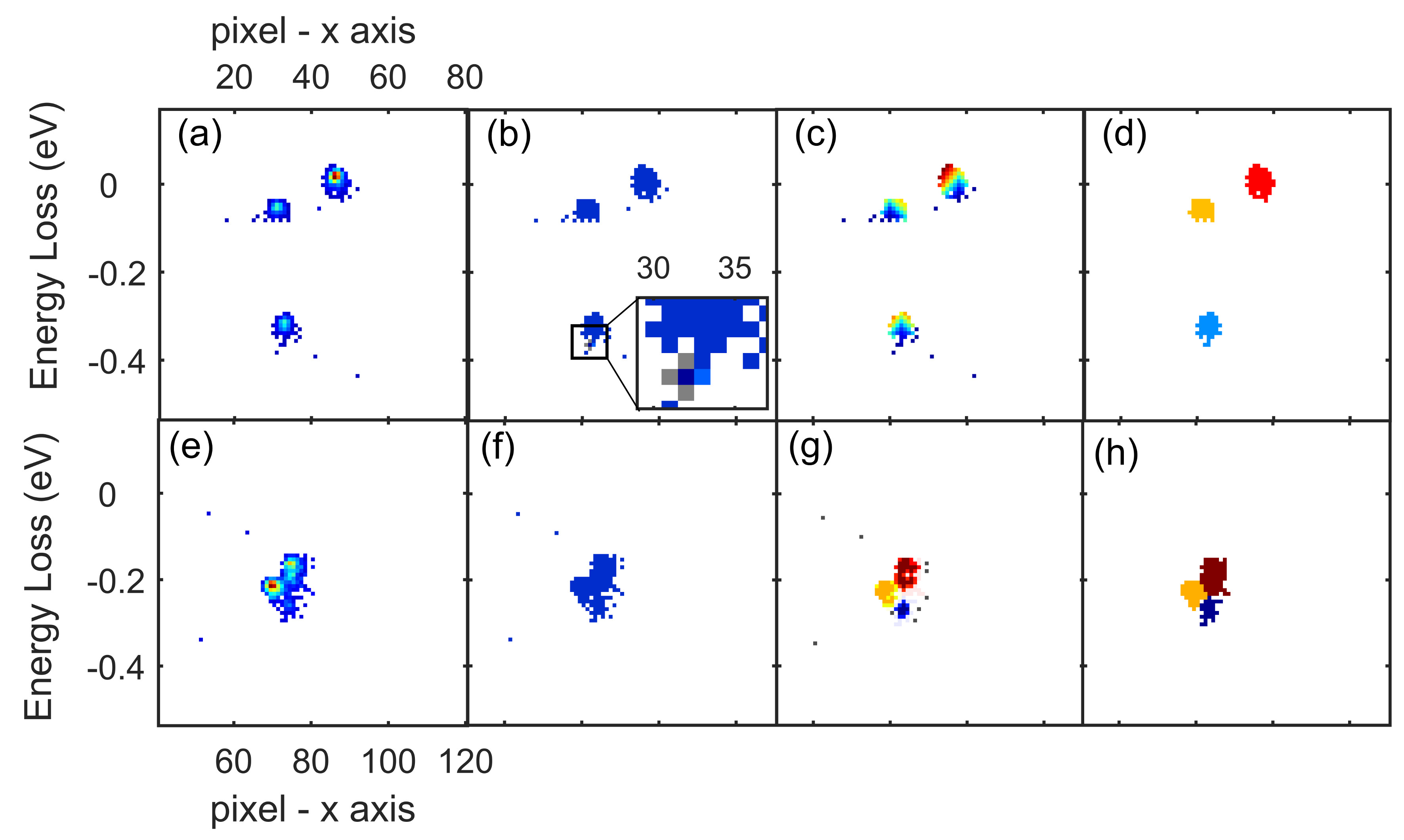}
\begin{singlespace}
\caption[Diagram.]{\label{fig:final_image.png}\small{Panel \emph{a} displays three candidate photon events from a single shot denoised raw image. Spurious single pixel spots are also visible. (b) The detection of connected regions follows a pixel connectivity criterion that is based on a Von Neumann neighbourhood (4-connected). The 3x3 square template used for the blob segmentation is shown in the inset, where is superimposed to a connected pixel. Darker hues indicate the  neighbours of the pixel being processed (c) Each region is then partitioned by using different labels (x$_i$,y$_j$). The sequential partition process is represented in false-colors from the bluish head to the reddish tail. (\emph{d}) Following the labeling stage, the image is partitioned into subsets of pixels. Spurious events, like noise spikes and hot spots whose connected area are less then few pixels ($\leq$4), are not considered. Panel (\emph{e}) displays a typical case of a multi-blobs event. Panel \emph{f-g-h} the labelling and partitioning processes is applied to multi-blobs events.}}
\end{singlespace}

\end{center}

\end{figure*}

\begin{figure*}[ht!]
\captionsetup{justification=centerlast}
\begin{center}

\includegraphics[width=0.8\textwidth]{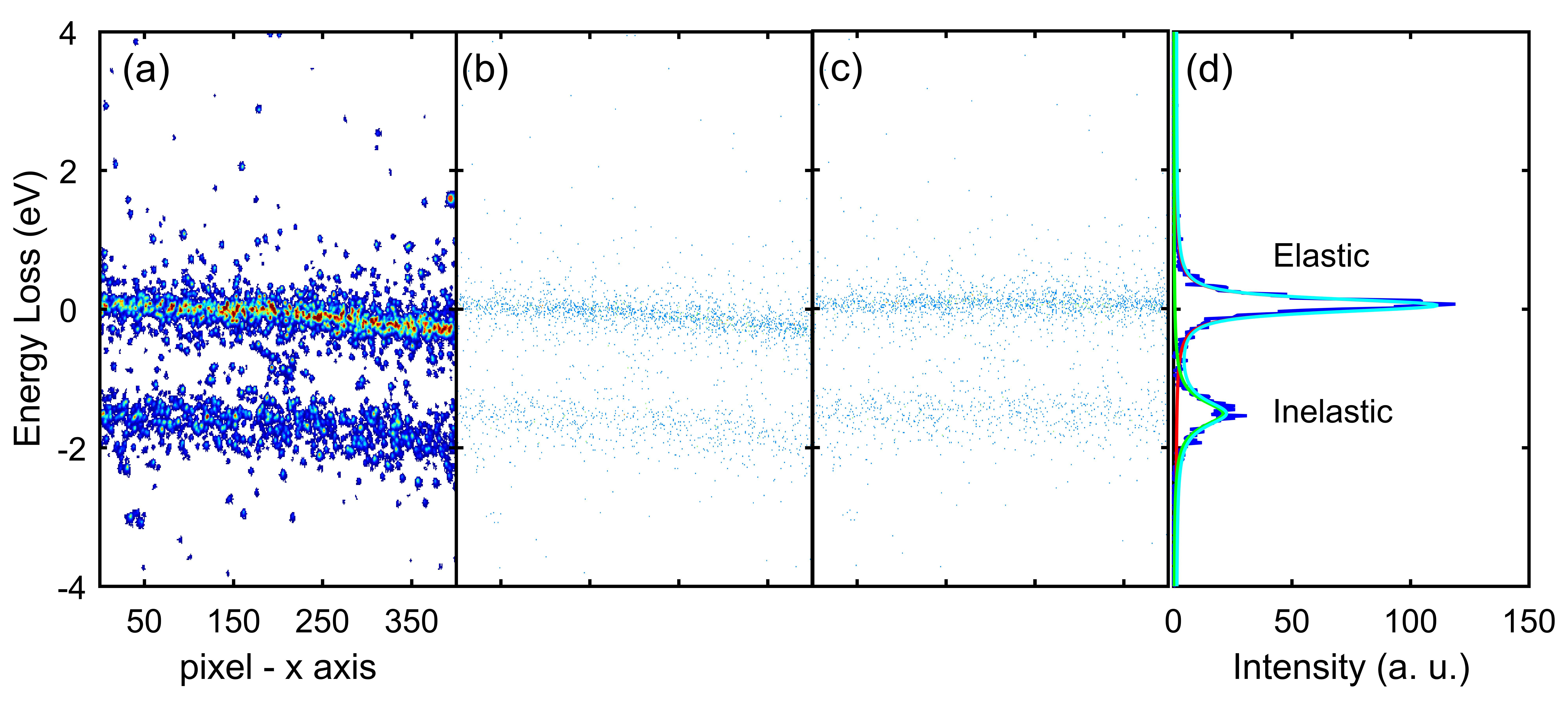}
\begin{singlespace}
\caption[Diagram.]{\label{fig:eighth_image.png}\small{Panel \emph{a}: an integrated raw RIXS image obtained by summing over $\sim$ 3000 single shot images ($\sim$60 seconds at 50 Hz FEL repetition rate). \emph{b}: the same 2D RIXS image reconstructed after the image recognition process for the exctraction of photon events. \emph{c}: correction of the parabolic coma originating from the VLS grating. \emph{d}: the integrated RIXS lineshape obtained integrating along the isoenergetic lines.}}
\end{singlespace}

\end{center}

\end{figure*}
\subsection{Noise reduction and image segmentation}

The single photon counting multi-hit detection device suffers from different sources of noise. 
One source of noise is the intrinsic stochastic nature of the electron multiplication process that leads to an unavoidable uncertainty of the MCP gain factor. This is responsible for an unpredictable variation of the intensity and size of the luminous clouds, which can saturate the MCP response and deteriorate the energy resolution.
In addition, the CMOS camera is also affected by statistical shot-noise (square root of counts dependence), dark current noise (3-4 counts per pixel at room temperature), leading to possible hot spots where pixels have a higher dark current. Finally, CMOS electronic noise originates from either the charge leakage among neighbouring pixels close to the point of saturation and from conversion errors produced in the amplifying circuit.

All these noise sources contribute to a background noise on the image that that needs to be removed for an efficient photon recognition processing.
Short exposure times, which are typical with a pulsed FEL source, certainly contribute to keep this background offset quite low.
 
The main aim of the image filtering is to achieve both noise reduction and feature preservation.
Noise reduction is achieved by image segmentation, which separates the background digital noise from detected photons.

Image segmentation consists of partitioning each single shot RIXS image into set of pixels (macropixels) by applying a thresholding method. 
A counts threshold defines the noise intensity level, above which macropixels represents photon events. Thus, each single shot RIXS frame is partitioned by subtracting a noise background.  
A segmented single shot RIXS frame is shown in Fig. \ref{fig:first_image.png}b, where photon events (or macropixels) are clearly visible together with spurious single pixel \texttt{hot spots}.

\subsection{Image deblurring}

Imaged photon traces may be affected by shape distortions originating by CMOS camera discharges, bad focusing etc. An example of is shown in Fig. \ref{fig:seventh_image.png}a. 
A de-blurring procedure is then applied before calculating the position of the centre of mass of each blob.
This procedure consists on an analytical refocusing obtained by an exponential rescaling of the pixel intensities (\texttt{k=log(max\_val)/max\_val}), while the maximum intensity (\texttt{max\_val}) of the blob is kept fixed.

The optical artefacts can in particular deteriorate the overall energy resolution because the calculated centre of mass of the distorted blob in general do not coincides with the position of the maximum intensity. 
Fig. \ref{fig:seventh_image.png}c clearly shows the cases where the position of the maximum intensity does not overlap with the centre of mass of the blob.

\subsection{Connected-area labelling and blob extraction}

The de-noising procedure enables an appropriate segmentation of the entire image by locating the photon event regions and by determining their boundaries. Fig. \ref{fig:final_image.png}a displays an example of image segmentation with three photon blob candidates. 

In order to recognize photon events and to distinguish them from spurious artefacts, each macro-pixel within a segmented image must satisfy a cluster connectivity criterion. In addition, the connected areas must have null intersection.
 
In order to identify connected areas, each pixel within the de-noised RIXS image, whose intensities are above the noise threshold, are sequentially scanned using a neighbourhood template, which maps and label each isolated blob. The connectivity check procedure is displayed in fig. \ref{fig:final_image.png}b-c.

The connected-component labelling process implemented in our algorithm follows a 4-connected connectivity criterion, where a pixel is connected to every pixel that touches one of its edges. An example of the neighbourhood template is displayed in the inset of Fig. \ref{fig:final_image.png}b and it is represented as a grey cross that spatially overlaps a single pixel.
This type of pixel connectivity is defined by a von Neumann neighbourhood:

\begin{equation*}
S = \{ (x_0+i,y_0+j) \in \mathds{N}^2 \ \big| \ i, j = 1, 0, -1 \ \wedge \ |i + j| = 1 \}
\end{equation*}

Then each connected region is partitioned by using different labels following a progressive path. 
In Fig. \ref{fig:final_image.png}c the progression of the partition process is represented in false-colors.
Finally, each labelled blob (see Fig. \ref{fig:final_image.png}d) is identified by its center of mass (x$_i$,y$_j$) - with x the pixel position and y the energy loss in eV.
Connected regions whose pixel dimension is below a fixed cutoff ($\leq$4 pixels) are not considered as blobs and are discarded. This filters out very small macropixels whose intensity is inconsistent with the experimental process.

Multi-blobs regions (see Fig. \ref{fig:final_image.png}e-h) may appear as a result of partially overlapped blobs in the raw image. These regions are further segmented in order to discern single events from multiple occurrences.

Finally, the center of mass of each photon spot is calculated and localized on the pixel matrix and it is stored together with the relevant FEL single shot parameters (e.g. FEL pulse intensity, photon wavelength, pulse shape, etc).

\subsection{reconstruction of the integrated RIXS spectrum}

A Cu M$_3$-edge RIXS raw image taken from a GeCuO$_3$ sample, integrated over $\sim$3000 FEL pulses 
and with the noise background removed, is shown in Fig. \ref{fig:eighth_image.png}, panel \emph{a}. 
The same image after the single photon retrieval method is shown in panel \emph{b} of the same figure.

The isoenergetic lines of the elastic/inelastic features are clearly visible.
The energy position of the elastic line is kept around the zero energy loss (0 eV). 
The typical inelastic peak, which originates from the inelastic scattering between the photons and the local orbital excitation of CuGeO$_3$, locates at an energy loss around -2 eV.
In order to obtain the RIXS spectrum, the RIXS image must be integrated along the isoenergetic lines.
The isoenergetic lines are not perfectly aligned because a misalignment of the detector.
In addition, the optical coma by the VLS grating is also present.
These optical aberrations are removed by a vertical shift correction $\Delta_c$, which is a second order function

\begin{equation*}
\Delta_c=a\cdot c+b\cdot c^2
\end{equation*}

fitting the locus of the elastic maxima and whose parameters depend only on the position (c column) of the pixel and on the photon energy.
The column shift applied to the matrix is an integer number of pixels.
Finally, the integrated RIXS lineshape (Fig. \ref{fig:final_image.png}d) is obtained by projecting the RIXS image along the x dimension.
 
\section{Conclusions}

RIXS spectroscopy can be extended to pulsed soft X-ray sources like free electron lasers with clear benefits like the possibility to carry out pump-probe and time dependent RIXS experiments.
In addition, this opens up the opportunity to acquire shot-to-shot RIXS spectra following the pulsed structure of the FELs source.

In this paper, we discuss the implementation of an analytical photon retrieval method for extracting number, energy and positions of the photons from single shot RIXS images.

\section{Acknowledgements}

The FERMI project at Elettra Sincrotrone Trieste is
supported by MIUR under Grants No. FIRB RBAP045JF2 and No. FIRB RBAP06AWK3.

\newpage
\providecommand{\newblock}{}

\end{document}